\documentclass[12pt,sd,amsmath,amssymb,reprint,superscriptaddress,floatfix,preprintnumbers]{iopart}
\usepackage{graphicx}% Include figure files
\usepackage{dcolumn}% Align table columns on decimal point
\usepackage{bm}% bold math

\usepackage[dvipsnames]{xcolor}
\usepackage{epstopdf}
\usepackage{hyperref}
\usepackage{iopams}
\usepackage{framed}
\usepackage{inputenc} %tableofcontents

\expandafter\let\csname equation*\endcsname\relax
\expandafter\let\csname endequation*\endcsname\relax
\usepackage{amsmath,amsfonts,amssymb,cancel}

\begin{document}

%\title[Author guidelines for IOP Publishing journals in  \LaTeXe]{How to prepare and submit an article for 
%publication in an IOP Publishing journal using \LaTeXe}
\title{Entropy production and fluctuation theorems on complex networks}

%\author{Content \& Services Team}
\author{Jaewoo Jung$^1$, Jaegon Um$^{1,2}$, Deokjae Lee$^1$, Yong W. Kim$^{3}$, D. Y. Lee$^{4}$, H. K. Pak$^{4,5}$, and B. Kahng$^1$}
%\address{IOP Publishing, Temple Circus, Temple Way, Bristol BS1 6HG, UK}
\address{$^1$CCSS, CTP and Department of Physics and Astronomy, Seoul National University, Seoul 08826, Korea \\
$^2$ BK21PLUS Physics Division, Pohang University of Science and Technology, Pohang 37673, Korea \\
$^3$ Department of Physics, Lehigh University, Bethlehem, PA 18015, USA \\
$^4$ Center for Soft and Living Matter, Institute for Basic Science, Ulsan 44919, Korea \\
$^5$ Department of Physics, Ulsan National Institute of Science and Technology, Ulsan 44949, Korea}
%\ead{submissions@iop.org}
\ead{dotoa@snu.ac.kr, sonswoo@hanyang.ac.kr, bkahng@snu.ac.kr}
\vspace{10pt}

\begin{abstract}
Entropy production (EP) is known as a fundamental quantity for measuring the irreversibility of processes in thermal equilibrium and states far from equilibrium. In stochastic thermodynamics, the EP becomes more visible in terms of the probability density functions of the trajectories of a particle in the state space. Inspired by a previous result that complex networks can serve as state spaces, we consider a data packet transport problem on complex networks. EP is generated owing to the complexity of pathways as the packet travels back and forth between two nodes along the same pathway. The total EPs are exactly enumerated along all possible shortest paths between every pair of nodes, and the functional form of the EP distribution is determined by extreme value analysis. The asymptote of the accumulated EP distribution is found to follow the Gumbel distribution. 
\end{abstract}

\section{Introduction}
The concept of entropy production (EP) has received increasing attention recently as nonequilibrium phenomena have become a central issue in statistical physics~\cite{Seifert2012,information,infor_thermo,jkps}. 
The fluctuation theorem (FT) of EP in the nonequilibrium steady state was established in Refs.~\cite{Evans1993,Evans1994,Gallavotti1995}. Crooks~\cite{Crooks1998}, Jarzynski~\cite{Jarzynski1996}, and others developed the FT for the dissipated work associated with other physical quantities such as the free energy. After the FT was first proposed for thermal systems, further studies were performed to obtain more general FTs and deeper understanding~\cite{Kurchan1998,Lebowitz1999}. As a result, EP could be viewed microscopically in terms of the trajectories of a single particle ~\cite{Seifert2005,esposito}. The EP was defined as the logarithm of the ratio of the probabilities that a dynamic process proceeds in the forward and corresponding reverse directions between two states in a nonequilibrium system~\cite{Kurchan1998,esposito}. The EP distribution is formed as the integral of those EPs over all possible states and trajectories. FTs such as the integral FT and detailed FT were derived based on the EP distribution~\cite{Seifert2012}.  

Both the EP and FT are well established formally and have been experimentally tested in terms of the work distribution for various experimental setups~\cite{ciliberto1,ciliberto2}, including RNA folding~\cite{rna}, colloidal suspensions~\cite{colloidal} and electric circuit~\cite{circuit}. However, the asymptotic functional form of the EP distribution has rarely been obtained explicitly, because it is extremely difficult  to experimentally find the probability density function of each trajectory along which a particle proceeds in the forward and its time-reverse directions~\cite{ciliberto2}. Moreover the number of trajectories increases exponentially as the number of steps is increased.

Here we recall that complex networks can serve as state spaces. For instance, each node in a protein folding network represents a protein conformation, and two nodes are connected by a link when a protein conformation is changed to another in consecutive steps~\cite{pin}. Recent studies using molecular dynamics simulations revealed that in protein folding dynamics, there exist a few major pathways involving multiple folds from denatured states to the natural state~\cite{pin,de_los_rios}, in disagreement with Levinthal's perspective~\cite{levinthal}. Rather it implies that the folding dynamics may proceed as biased random walks along the shortest pathway on the conformation network. 

Inspired by this previous research, in this paper, we consider a biased random walk problem on complex networks. This biased random walk problem is closely related to a data packet transport problem on Internet in free flow state, a well known problem in network science community. Suppose that a data packet is sent from one node to another on a complex network such as the Internet. At each time step, the packet is transmitted to a neighbor according to the router protocol at each node toward the final destination. Unless traffic is congested, packets generally travel along the shortest path between a starting node and a final destination. Thus, to consider only the shortest pathways is quite natural for dynamics on complex networks. Every possible shortest pathway between any pair of nodes can be identified within the computational complexity $O(N^2\log N)$, where $N$ is the network size~\cite{Dijkstra1959,newman_btw1,newman_btw2}. On the other hand, the flow along the shortest pathways on complex networks was used to quantify a person's influence in society~\cite{newman_btw1} and the load on a router on the Internet~\cite{Goh2001}. 

Owing to topological diversity of the shortest pathways, the probability that a packet  takes one shortest pathway from one node to the other in forward direction cannot be the same as that in reverse direction on the same pathway, which generates nonzero EP. We collect all EPs obtained from every possible shortest pathway of each pair of nodes. Thus the dataset is complete. The EP distribution consists of $N_{\rm EP}\equiv n_{\rm sp}N(N-1)$ values, where $n_{\rm sp}$ is the mean number of shortest pathways between a pair of nodes.

Next, we perform extreme value (EV) analysis~\cite{gumbel} to determine the functional form of the asymptotic behavior of the EP distribution, called an asymptote. This method was originally developed to predict the probability of rare events such as floods of a certain level, for instance. To perform this EV analysis, we use the theorem that states that the functional form of the asymptote of a distribution is related to the distribution of the maxima of samples selected randomly from the original $N_{\rm EP}$ elements~\cite{gumbel}. Using this property, we find that the Gumbel distribution is the best fit to the asymptote of the $n$-th power of the accumulated EP distribution. Moreover, the Gumbel distribution seems to be common to other similar distributions obtained from different complex networks. On the basis of this result, we assert that the EP distribution behaves asymptotically as the Gumbel distribution.

Networks are a platform for interdisciplinary studies. The nodes and links of a network represent routers and optical cables on the Internet, web documents and hyperlinks on the World Wide Web, and individuals and social interactions in social networks~\cite{netsci_book}. It was found that most complex networks in the real world are heterogeneous in the number of connections at each node, called the degree. Their degree distributions follow a power-law or heavy-tailed distribution. Because the exponent $\lambda$ of the degree distribution $P_d(k)\sim k^{-\lambda}$ is in the range $2< \lambda \le 3$ in many real-world systems, they are called scale-free networks. On the other hand, a random network introduced by Erd\H{o}s and R\'enyi (ER) has a degree distribution following the Poisson distribution. We have considered the data packet transport problem on diverse types of model networks and find that the Gumbel distribution seems to fit all scale-free networks well. 

This paper is organized as follows. In Sec.~\ref{sec:ep}, we introduce an EP induced by the topological complexity of the shortest pathways on networks. In Sec.~\ref{sec:ft}, we show that the total EP obtained from every shortest pathway satisfies the integral FT and the detailed FT. In Sec.~\ref{sec:numerical}, we obtain the EP distribution for several model networks numerically. In Sec.~\ref{sec:asymptotes}, we determine the functional type of the asymptote of the accumulated EP distribution using the EV approach. We confirm that the asymptote follows the Gumbel distribution. A summary is presented in Sec.~\ref{sec:summary}.

\section{EP on networks}\label{sec:ep}

\begin{figure}[!t]
\includegraphics[width=\linewidth]{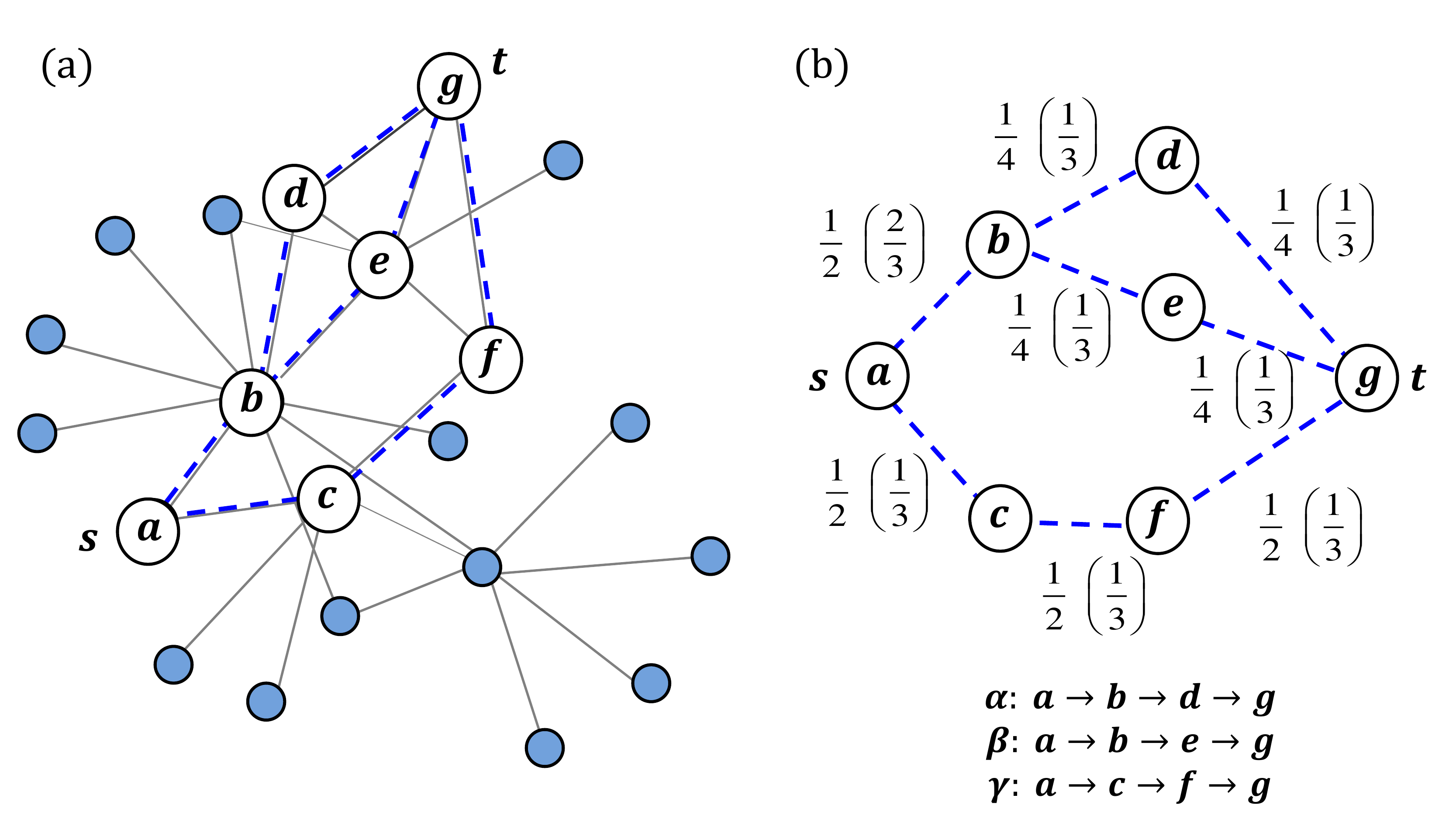}
\caption{(a) Sample network to illustrate the EPs along each shortest path from $a$ to $g$ and shortest return path from $g$ to $a$. For pathway $\alpha$, a data packet starts along the pathway $a\to b\to d\to g$ and returns in the reverse direction. (b) At node $a$, there are two ways to move toward node $g$ with equal probability. The packet takes the link $a\to b$ with probability $1/2$. Next, it takes the link $b\to d$ with probability $1/2$. The link $d\to g$ is taken with probability one. Accordingly, the transition probability along the pathway $\vec{\alpha}$, denoted as $\Pi[\vec{\alpha};a, g]$, is found to be $1/4$. In the reverse trajectory $\vec{\alpha}'$, the transition probability $\Pi[\vec{\alpha}'; g,a]$ is found to be $1/3$. Table~\ref{Example of the EPs} shows the transition probabilities along each shortest pathway in the forward and corresponding reverse directions.}
\label{fig:sample}
\end{figure}

We consider data packet transport from a source node $i$ to a target node $j$ on a given network along a shortest pathway $\vec{\alpha}$ of length $d_{st}$, where $\vec{\alpha} = (\alpha_0, \alpha_1, \cdots, \alpha_{d_{st}-1}, \alpha_{d_{st}})$. In the sequence $\vec{\alpha}$, each element stands for the node on the shortest pathway with the boundary condition, $\alpha_0=i$ and $\alpha_{d_{st}}=j$. Then the probability $P[\vec{\alpha}]$ that transport occurs along the pathway $\vec{\alpha}$ is given as $P[\vec{\alpha}] = \rho(i) \rho(j|i) \,\Pi[\vec{\alpha} ; i,j] $, where $\rho(i)$ denotes the probability that node $i$ is selected as a source and $\rho(j|i)$ is the conditional probability that node $j$ is chosen as a target. Given the pair of source and target, $(i,j)$, the transition probability $\Pi[\vec{\alpha}; i,j]$ along $\vec{\alpha}$ is determined by the topology of the shortest pathways from $i$ to $j$. The shortest pathway is either single or multiple, and multiple pathways are either in parallel or entangled, as shown in Fig.~\ref{fig:sample}. As a packet travels along the shortest pathways, it can reach a branching node. Then, the packet chooses one branch among all the branches with probability $1/N_b$, where $N_b$ is the number of branches on the shortest pathway. This random choice is repeated as the packet reaches a branching node. When the packet reaches a target, the probability of taking that shortest pathway can be calculated as the product of those probabilities, as illustrated in Fig.~\ref{fig:sample}.
The effect of this random choice mimics the stochastic noise in dynamic process. Let us consider the reverse process where the packet returns along the corresponding reverse path $\vec{\alpha}'$ of $\vec{\alpha}$ from node $j$ to node $i$, where $\vec{\alpha}' = (\alpha_{d_{st}}, \alpha_{d_{st}-1}, \cdots, \alpha_1, \alpha_0)$. Since $\vec{\alpha}'$ is also one of the shortest pathways from the source $j$ to the target $i$, one can define the probability $P[\vec{\alpha}']$ in the same manner as $P[\vec{\alpha}]$.  
In general, the probabilities for the path $\vec{\alpha}$ and its reverse one $\vec{\alpha}'$  
may be different. The discrepancy can be regarded as the irreversibility for the transport along
$\vec{\alpha}$, and thus the corresponding EP is defined as follows: 
\begin{align}
\Delta S[\vec{\alpha}] =\ln \Big [\frac{P[\vec{\alpha}]}{P[{\vec{\alpha}'}]}\Big] 
=\ln \Big[\frac{\rho(i)\rho(j|i)\,\Pi[\vec{\alpha}; i,j]} {\rho(j)\rho(i|j)\,\Pi[\vec{\alpha}'; j,i]}\Big]\,.
\label{Definition of the EP}
\end{align}

In this problem, $\rho(i)=\rho(j)=1/N$, because the node is selected randomly from among $N$ nodes. 
The conditional probability is also given by $\rho(j|i)=\rho(i|j)=1/(N-1)$ because the node $j$ ($i$) is randomly selected from $N-1$ nodes excluding node $i$ ($j$). Therefore, the non-zero EP in Eq.~\eqref{Definition of the EP} is caused by only the difference between $\Pi[\vec{\alpha} ;i,j]$ and $\Pi[\vec{\alpha}'; j, i]$. We will show that $\Pi[\vec{\alpha};i,j]$ can differ from $\Pi[\vec{\alpha}';j,i]$ owing to the topological diversity of the shortest pathways on complex networks.  

We consider a simple example to explain how to calculate the transition probabilities on the shortest pathways. Fig.~\ref{fig:sample} is a subgraph of a network showing the shortest pathways between two nodes, $a$ and $g$ as the source ($s$) and target ($t$), respectively. There exist three shortest pathways, which are denoted as $\vec{\alpha}$, $\vec{\beta}$, and $\vec{\gamma}$, with the length $d_{st}=3$. Let us first consider packet transport along the pathway $\vec{\alpha} = (a,b,d,g)$, from $a$ toward node $g$. At node $a$, the packet needs to choose either node $b$ or node $c$, which we assume are chosen with equal probability, as the site of the next step. Thus, hopping from $a$ to $b$ occurs with probability $1/2$, as does hopping from $a$ to $c$. Next, it chooses node $d$ with probability $1/2$, because the pathway is divided into two possibilities. 
Thus, the packet arrives at node $d$ with probability $1/4$. Then it travels to the target $t=g$ without any branching, i.e., with probability one. Accordingly, the transition probability is given as $\Pi[\vec{\alpha}; a,g]=a\xrightarrow{1/2}b\xrightarrow{1/2}d\xrightarrow{1} g=1/4$. On the other hand, when it returns from node $g$ to $a$ along the reverse trajectory $\vec{\alpha}' = (g,d,b,a)$, 
one can see $\Pi[\vec{\alpha}'];g,a]= g\xrightarrow{1/3}d\xrightarrow{1}b\xrightarrow{1} a=1/3$. Thus, the two transition probabilities are not the same: $\Pi[\vec{\alpha}; a,g] \ne \Pi[\vec{\alpha}';g,a]$. Further, $\rho(a)=\rho(g)=1/N$, and $\rho(j|i)=\rho(i|j)=1/(N-1)$, which yield $\Delta S[\vec{\alpha}] =\ln (3/4)$ by the definition Eq.~\eqref{Definition of the EP}. The EPs along the pathways $\vec{\beta}$ and $\vec{\gamma}$ can be similarly calculated and are listed in Table I. One can easily find that the transition probability is normalized as $\Pi[\vec{\alpha};a,g] +\Pi[\vec{\beta};a,g] + \Pi[\vec{\gamma};a,g] =1$. Therfore, for all possible shortest pathways of $N(N-1)$ pairs on the complex network, the probability $P[\vec{\alpha}]$ is also normalized, $\sum_{\vec{\alpha}} P[\vec{\alpha}] =1$.  

\begin{table*}[!ht]
\centering
\begin{tabular}{ c  c  c  c  c c} \\ 
\hline
\\ [0.01pt]
~~Pathway~~  & ~~Pathway from $s$ to $t$~~ & ~~~~$\Pi$~~~~ & ~~Pathway from $t$ to $s$ ~~ & ~~~~$\Pi$~~~~ & ~~~$\Delta S$~~~ \\ [5pt]		\hline
\hline 
\\ [0.01pt]
$\vec{\alpha}$ & $a$ $\xrightarrow{1/2}$ $b$ $\xrightarrow{1/2}$ $d$ $\xrightarrow{1}$ $g$ & $\frac{1}{4}$ & $g$ $\xrightarrow{1/3}$ $d$ $\xrightarrow{1}$ $b$ $\xrightarrow{1}$ $a$ & $\frac{1}{3}$ & $\ln \frac{3}{4}$ \\ [5pt]
$\vec{\beta}$ & $a$ $\xrightarrow{1/2}$ $b$ $\xrightarrow{1/2}$ $e$ $\xrightarrow{1}$ $g$ & $\frac{1}{4}$ & $g$ $\xrightarrow{1/3}$ $e$ $\xrightarrow{1}$ $b$ $\xrightarrow{1}$ $a$ & $\frac{1}{3}$ & $\ln \frac{3}{4}$ \\ [5pt]
$\vec{\gamma}$ & $a$ $\xrightarrow{1/2}$ $c$ $\xrightarrow{1}$ $f$ $\xrightarrow{1}$ $g$ & $\frac{1}{2}$ & $g$ $\xrightarrow{1/3}$ $f$ $\xrightarrow{1}$ $c$ $\xrightarrow{1}$ $a$ & $\frac{1}{3}$ & $\ln \frac{3}{2}$\\ [5pt]
\hline 
\end{tabular}	
\caption{Probability that a packet takes each shortest pathway and corresponding entropy production. $\Pi$ denotes the transition probability.}
\label{Example of the EPs}
\end{table*}

\section{Fluctuation theorems}\label{sec:ft}

Here we obtain the EP distribution over all possible shortest pathways between every pair of nodes. 
The EP distribution $P(\Delta S)$ is given by
\begin{eqnarray}
P(\Delta S) 
&=&\sum_{\vec{\alpha}}\delta(\Delta S-\Delta S[\vec{\alpha}]) \rho(i)\rho(j|i)\,\Pi[\vec{\alpha};i,j] \nonumber \\
&=&\sum_{\vec{\alpha}'}\delta(\Delta S-\Delta S[\vec{\alpha}]) \rho(j)\rho(i|j)\,\Pi[\vec{\alpha}';j,i]e^{\Delta S[\vec{\alpha}]}\nonumber \\
&=&\sum_{\vec{\alpha}'} \delta(\Delta S+\Delta S[\vec{\alpha}'] ) \rho(j)\rho(i|j)\,\Pi[\vec{\alpha}';j,i] e^{-\Delta S[\vec{\alpha}']}\nonumber \\
&=&P(-\Delta S)e^{\Delta S}\,, \label{eq:2}
\end{eqnarray}
where we have used the fact that $\Delta S[\vec{\alpha}']=-\Delta S[\vec{\alpha}]$ and $\sum_{\vec{\alpha}}$ can be replaced by $\sum_{\vec{\alpha}'}$ because the Jacobian is $1$. The relation $P(\Delta S)=P(-\Delta S)e^{\Delta S}$ is known as the detailed FT and is an instance of the Gallavotti--Cohen symmetry of the probability density function~\cite{Gallavotti1995}. From Eq.~\eqref{eq:2}, the integral FT is derived as $\langle e^{-\Delta S} \rangle = \sum_{\Delta S}  e^{-\Delta S} P(\Delta S) =1$. 

\begin{figure}[!ht]
\centering
\includegraphics[width=0.6\linewidth]{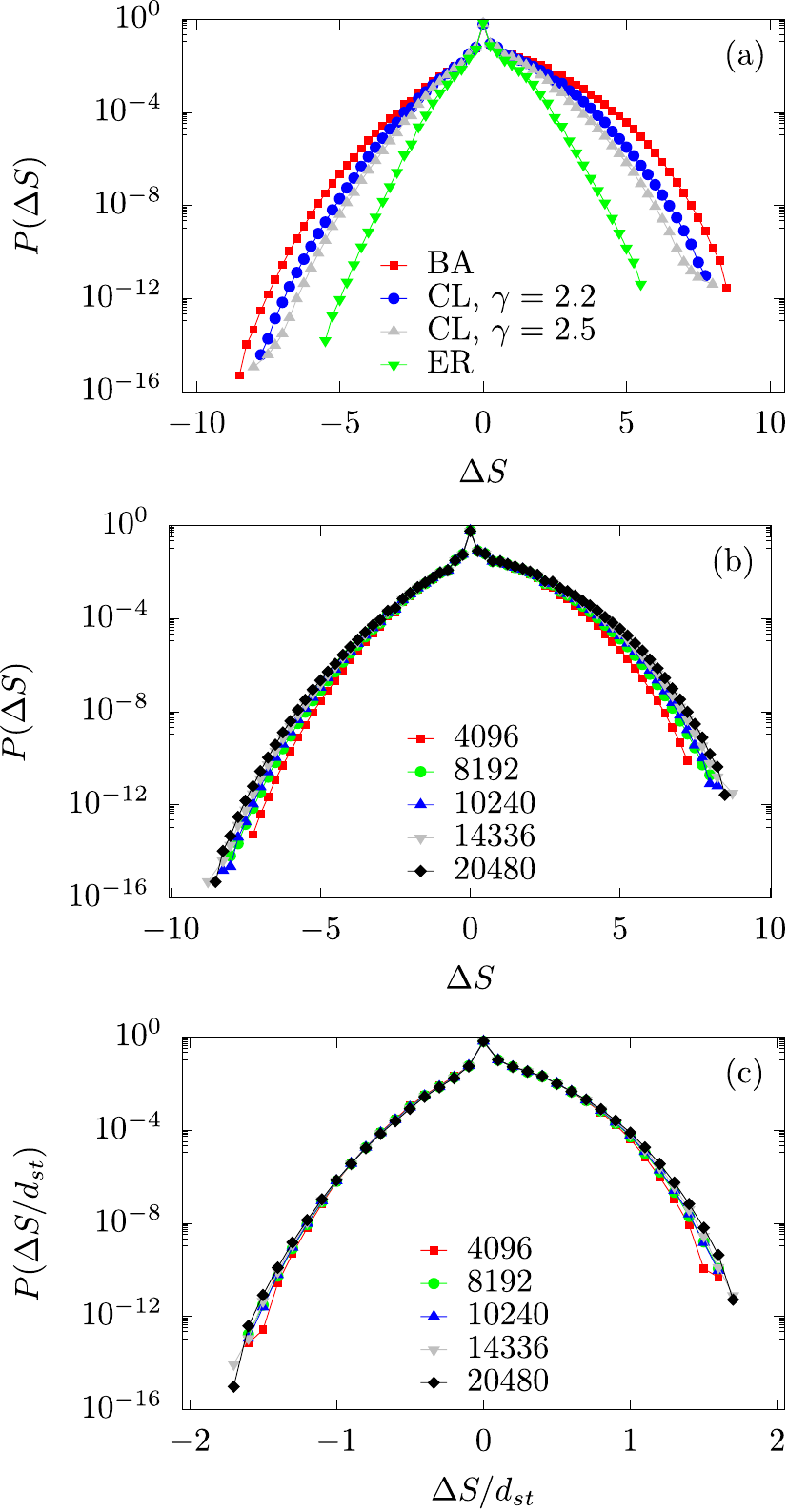}
\caption{(a) EP distributions on the four model networks: BA model, scale-free CL model with degree exponent $\gamma=2.2$, scale-free CL model with degree exponent $\gamma=2.5$, and ER model, from top to bottom. Data are obtained from the giant component of each model network of system size $N=2^{11}\times 10$ and mean degree $\langle k \rangle=8$. They are averaged over 300 configurations. All EP distributions exhibit peaks at $\Delta S=0$, which are attributed to transport along untangled pathways. (b) EP distribution on BA networks of different system sizes, $N=4096, 8192, 10240, 14336$, and 20480. As $N$ is increased, the EP curves tend to converge to the asymptotic one. (c) Distribution of EPs divided by the Hamming distance $d_{st}$ between a source (s) and a target (t) for each pathway, that is, $\Delta S/d_{st}$. The system sizes are the same as those in (b). The data for the different system sizes collapse onto a single curve.}
\label{fig:ep_dist}
\end{figure}

\section{Numerical results}\label{sec:numerical}

We perform numerical simulations to obtain EPs based on transport along every shortest pathway between all possible pairs of nodes on several networks: the Barab\'asi--Albert (BA) model~\cite{ba}, ER model~\cite{er}, and Chung--Lu (CL) model~\cite{cl} with the degree exponents $\gamma$ = 2.2 and $\gamma=2.5$. We will explain these three model networks in Appendix. 
The EP distributions $P(\Delta S)$ on these networks are shown in Fig.~\ref{fig:ep_dist}(a). All these networks were constructed with the same mean degree $\langle k \rangle=8$ and system size $N=2^{11}\times 10$. We obtain these EP distributions on the giant component of each network. The EP distributions have different shapes. The width of the EP distribution on the BA model is generally wide, whereas that on the ER model is generally narrow. This result arises from the extent of the topological diversity of each type of network.  

In statistical mechanics, the entropy is an extensive quantity with respect to the system size $N$. However, in this problem, the length $d_{st}$ of each pathway plays a role similar to that of $N$ in Euclidean space. Thus, we rescale the EP $\Delta S$ by the path length and define $\Delta S/d_{st}$. The EP distributions obtained for different network sizes $N$ collapse onto a single curve, as shown in Figs.~\ref{fig:ep_dist}(b) and \ref{fig:ep_dist}(c). 
%\xnew{Therefore, $\delta s \equiv \Delta S/d_{st}$ plays a role of the large deviation function of the entropy production.}  

\begin{figure}[!h]
\centering
\includegraphics[width=0.6\linewidth]{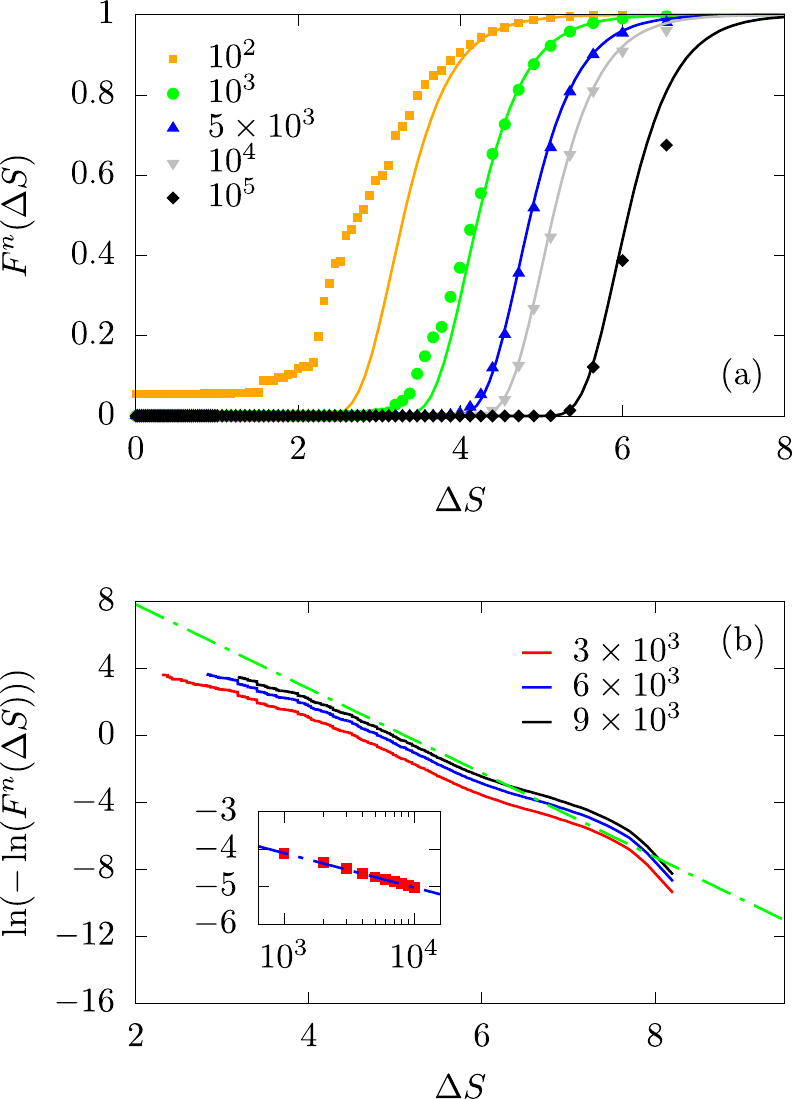}
\caption{(a) Test of the Gumbel distribution for the $n=(\ell m)$-th power of the accumulated EP distribution, $F^{n}(\Delta S)$, which show for BA model networks of size $N/10=2^{11}$. Data points (symbols) are obtained by exact enumeration. Solid curves represent $F^{\ell}(a_m\Delta S+b_m)$ with a fixed $\ell=1000$ but with different $m$ values, making $n(=\ell m)=10^2, 10^3, 5\times 10^3, 10^4$, and $10^5$ from left to right. The solids curves are parallel to the right and shifted depending on $m$. This pattern implies $a_{m}=1$, and $b_{m}\propto \ln m$. One can see that for $n=10^3-10^4$, solid curves seem to be fit to the data points to some extent. (b) Plot of $\ln(-\ln(F^{n}(\Delta S)))$ versus $\Delta S$ to test $(1/c)(\Delta S-c\log_{10}m)$ for $\ell=10^3$ and $m=3$, $6$, and $9$. Parallel alignment of data points for different values of $m$ to the straight dash-dotted line implies $a_m=1$. The dash-dotted line is a guideline with a slope $-2.52$ obtained by taking $1/c$, where $c\approx 0.397$ is measured by the plot of $b_{m}$ versus $\log_{10} m$ in the inset.}
\label{fig:gumbel}
\end{figure}

\section{Asymptotes of the EP distribution}\label{sec:asymptotes}

It is interesting to determine the functional form of the asymptotic EP distribution, called asymptote, because the mean entropy production is related to the large deviation function~\cite{large_deviation}. We follow the Fisher and Tippett method for EV analysis~\cite{fisher} to perform this task. We consider the dataset composed of $N_{\rm EP}$ EPs obtained from all possible shortest pathways between every pair of nodes of a given network such as the BA model, for instance. Next, we select $\ell$ elements randomly from among $N_{\rm EP}$ elements and construct a set. Repeating this construction $m$ times, we set up $m$ sets of size $\ell$. One can consider two asymptotes: one for positive and the other for negative tails. For the positive tail, let us consider another set $\{y_i\}$ composed of the largest elements $y_i$ of each set $i=1,\dots,m$. Then the largest value of the elements of the set $\{y_i\}$ ($i=1,\cdots, m$) is the largest value of those $\ell m$ elements selected randomly from $N_{\rm EP}$, where the elements are not necessarily distinct. To quantify this, we consider the accumulated distribution of $P(\Delta S)$, i.e., $F(x)={\rm Prob}\{\Delta S \le x\}$, that is, $F(x)=\int_{-\infty}^x d\Delta S P(\Delta S)$. The probability that the largest value $y_i$ of a set $i$ of size $\ell$ is less than $x$ is given as $F^{\ell}(x)$, which is denoted as $G_\ell(x)$ for later discussion. Next, the probability that the largest value among those $\{y_i\}$ ($i=1,\cdots, m$) elements is less than $x$ is given as $G_\ell^m(x)$, which is equal to $F^{\ell m}(x)$. If there exists the asymptote $G_\ell(x)$ of $F(x)$ for large $\ell$, $G_{\ell}(x)$ and $G_{\ell}^m(x)$ would have the same functional form. 
%At this stage, we recall the EV statistics theory. Let us define an asymptote $G(x)$ as
%\begin{align}
%G(x) &\equiv {\rm Prob}\bigg\{\frac{M_{\ell} - b_\ell}{a_\ell}\le x \bigg\}  \\
%&= {\rm Prob}\bigg\{M_{\ell} \le a_\ell x + b_\ell \bigg\}, 
%\label{Gdef}
%\end{align}
%where $M_\ell$ is a random variable representing the maximum value from a set of size $\ell$. $a_\ell$ and $b_\ell$ are appropriate sequences of $\ell$ to make the maximum value  $M_\ell$ bounded for general $\ell$. We consider $m$ set of size $\ell$, in which the maximum element $M_{\ell m}$ 
Because a linear transformation of $x$ does not change the form of the distribution, one may think that $G_\ell(x)$ satisfies the following stability postulate~\cite{gumbel}:
\begin{align}
F^{\ell m}=G_\ell^m(x)=G_\ell(a_m x+b_m).
\label{stability_postulate}
\end{align}

Here we introduce a function $G(x)$ that satisfies the relation $\lim_{\ell \to \infty}G_\ell(\alpha_\ell x+\beta_\ell)=G(x)$.  
It is known that there exist three types of functional forms for $G(x)$: i) Gumbel type, ii) Fr\'echet type, and iii) Weibull type. They have the following simplified forms: i) $G(x)={\rm exp}[-{\rm exp}(-x)]$ with $-\infty < x < \infty$, ii) $G(x)={\rm exp}[-x^{-\zeta}]$, where $\zeta$ is a constant and $0 < x < \infty$, and iii) $G(x)={\rm exp}[-(-x)^{\eta}]$, where $\eta$ is a constant and $-\infty < x < 0$. 
%\begin{align}
%F^\ell(x)=F(a_\ell x+b_\ell).
%\end{align}
%It is known that there exist three types of functional forms for $F(x)$: i) Gumbel type, ii) Fr\'echet type, and iii) Weibull type. 

We focus on the Gumbel distribution~\cite{gumbel} with $\alpha_\ell=1$, the relevant case to our problem. Then one can find easily that 
\begin{align}
G^n(x)=G(x-\beta_n)=G(x+b_n),
\label{eq:gumbel_am_1}
\end{align} 
where we replace the notation $-\beta_n$ by $b_n$ as used in Eq.~\eqref{stability_postulate}.
To determine $b_n\equiv b_{\ell m}$, we use $G^{\ell m}(x)=G^{\ell}(x+b_m)=G(x+b_{\ell m})$, so that $b_{\ell m}=b_\ell+b_m$. Thus, $b_m$ has the form $b_m=-c\ln m$ with a constant $c$. $c > 0$, because $G(x)$ is an increasing function with respect to $x$ and $G(x) < 1$, so that $G^n(x)$ is a decreasing function with respect to $n$. Eq.~\eqref{eq:gumbel_am_1} may be rewritten as 
\begin{align}
\ln n+\ln(-\ln G(x))=\ln(-\ln(G(x+b_n))).
\end{align}
Therefore, one can obtain 
\begin{align}
\ln(-\ln G(x))-\frac{x\ln n}{b_n}=r,
\end{align} 
where $r$ is independent of $x$. Using $b_n=-c\ln n$ ($n=\ell m$) with a positive constant $c$, one can obtain a functional form of $G(x)$ as 
\begin{equation}
G(x)=\exp(-e^{-\frac{1}{c}(x-cr)}).
\end{equation}

\begin{figure}[!h]
\centering
\includegraphics[width=\linewidth]{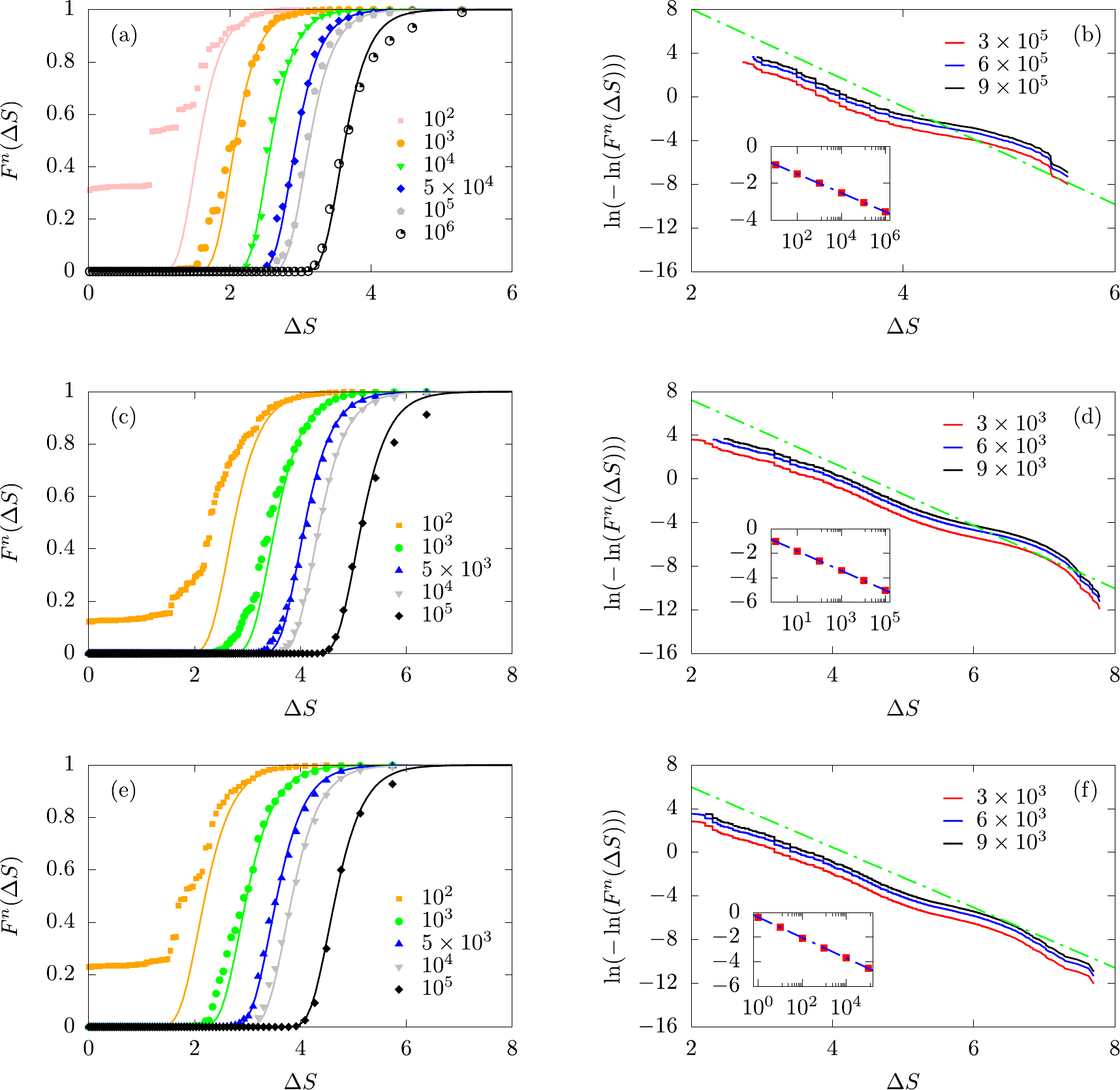}	
\caption{Similar plots to Figs.~\ref{fig:gumbel}(a) and \ref{fig:gumbel}(b) but on different networks, ER networks (a) and (b), and CL networks with degree exponent $\gamma=2.2$ (c) and (d) and with degree exponent $\gamma=2.5$ (e) and (f). The system sizes of all the networks are taken as $N/10 = 2^{11}$. Insets: Plots of $b_{m}$ versus $\log_{10} m$ to obtain $-c$. It is estimated that $c\approx 0.224$ (b), $0.347$ (d), and $0.362$ (f).}
\label{fig:gumbel_all}
\end{figure}

In Fig~\ref{fig:gumbel}, we find numerically that for a fixed $\ell=1000$, indeed $a_m=1$ as shown in Fig.~\ref{fig:gumbel}(a). The guide curves (solid curves) are parallel and offset to each other for different values of $m$. We also find in the inset of Fig~\ref{fig:gumbel}(b) that $b_m\approx -0.42\ln m$. Thus $c\approx 0.42$. These numerical results suggest that the functional form of $F^\ell(x)$ is of the Gumbel type, i.e., 
\begin{equation}
F^\ell(x)\equiv G_\ell(x)=\exp(-e^{-\frac{1}{c}(x-c\ln \ell-cr)})=\exp(-\ell e^{-x/c+r}).
\end{equation}
We note that in Fig.~\ref{fig:gumbel}(a), the data points for $n=10^3$, $5\times 10^3$, and $10^4$, where $n=\ell m$ seem to be fit to the theoretical curves $F^n(\Delta S)$. However, for $n=10^5$, the data points deviate from the curve. In fact, the value $n$ is comparable to the system size $N\simeq 2\times 10^5$. Because $\Delta S \sim d_{st}$ and the maximum separation between two nodes is bounded by $\sim \ln N$ in complex networks, the maximum EP, i.e., $\Delta S_{\rm max}$ is hard to be extended unless the system size is increased drastically in simulations. If we had extended the system size, we could have got more accurate data for larger $n$ values. On the other hand, similar plots to Figs.~\ref{fig:gumbel}(a) and \ref{fig:gumbel}(b) but on different model networks are shown in Fig.~\ref{fig:gumbel_all}. We find that $a_m=1$ and $b_m\propto \ln m$ are universal, but the constant $c$ depends on networks. We obtain $c=0.397, 0.224, 0.347$, and $0.362$ for BA networks, ER networks, and CL networks with degree $\gamma=2.2$ and $2.4$ from the insets of Figs.~\ref{fig:gumbel}(b) and ~\ref{fig:gumbel_all}(b), \ref{fig:gumbel_all}(d), and \ref{fig:gumbel_all}(f), respectively. The dash-dotted guide lines with slopes of $1/c=2.52, 4.47, 2.88$ and $2.76$ are drawn to compare the theoretical prediction to the empirical data in Figs.~\ref{fig:gumbel}(b) and ~\ref{fig:gumbel_all}(b), \ref{fig:gumbel_all}(d), and \ref{fig:gumbel_all}(f), respectively. Finally we note that we tried, without success, to fit the numerical data to other types of asymptotes ii) Fr\'echet type and iii) Weibull type.   

Once we found the functional type of $F^\ell(x)$, then the EP distribution is derived directly as
\begin{align}
P(\Delta S)=\frac{dF^\ell(x)}{dx}\frac{1}{{\ell F^{\ell-1}(x)}}\Big|_{x=\Delta S}
=\frac{1}{c}\exp(-e^{-x/c+r})e^{-x/c+r}\Big|_{x=\Delta S}
\to {\rm exp}(-\frac{1}{c}\Delta S)
\label{eq:9}
\end{align} 
in the limit $\Delta S \to \infty$. 

\begin{figure}[!h]
\centering
\includegraphics[width=\linewidth]{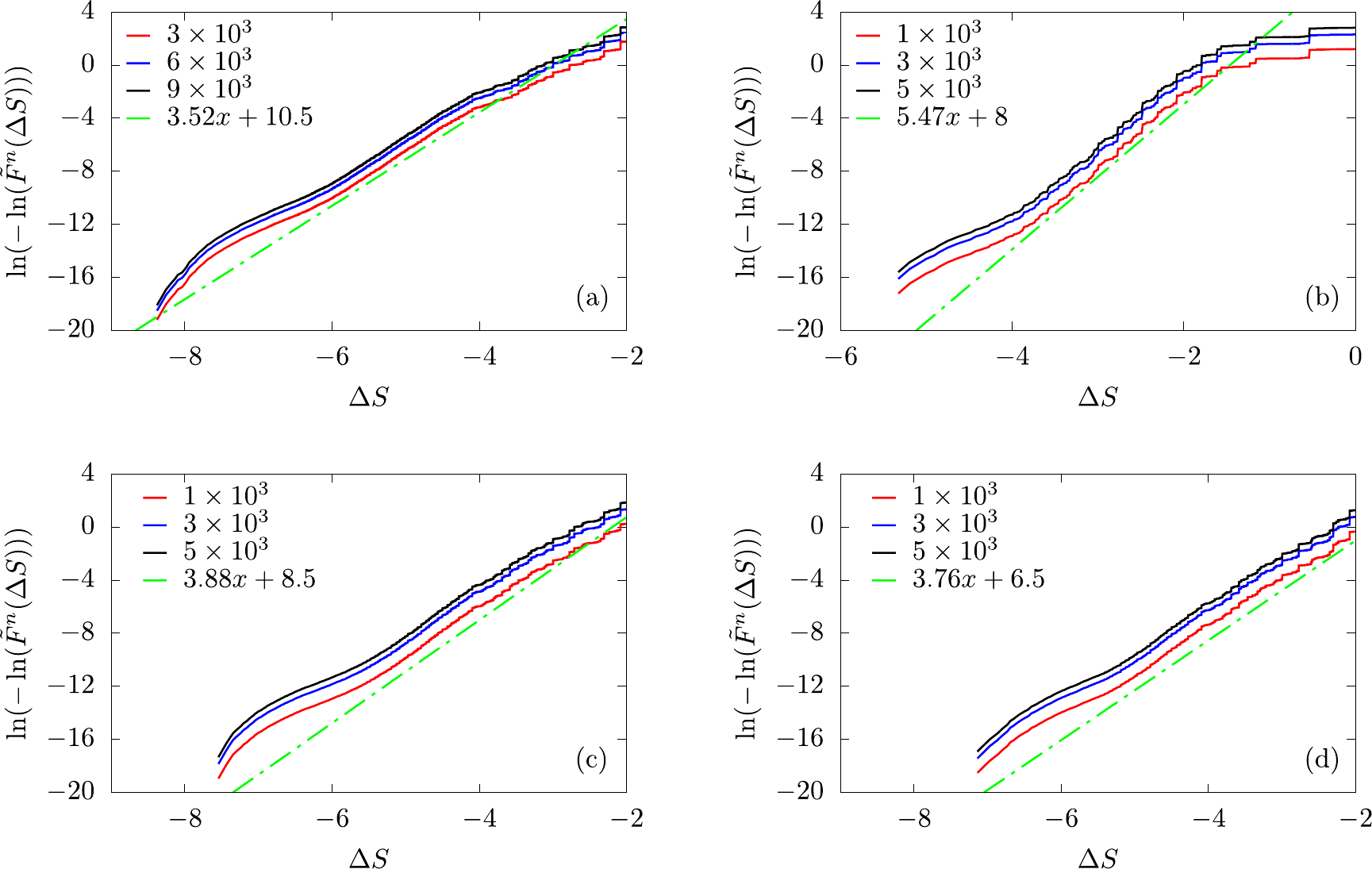}	
\caption{
Plot of $\ln(-\ln(\tilde F^{n}(\Delta S)))$ versus $\Delta S$ to check $({1}/{\tilde c})(\Delta S+\tilde c \log_{10}m)$ for $\ell=10^3$ and $m=3$, $6$, and $9$. $n=\ell m$. Parallel alignment of data points for different values of $m$ to the straight dash-dotted line implies $a_m=1$. The dash-dotted lines have slopes (a) $3.67$ for BA networks, (b) $6$ for ER networks, (c) $3.87$ for CL networks with $\gamma=2.2$, and (d) $3.87$ for CL networks with $
\gamma=2.5$. These values correspond to $1/\tilde c$. The system sizes of all the networks are taken as $N/10 = 2^{11}$.}
\label{fig:gumbel_negative}
\end{figure}

Next, we determine the functional form of the asymptote of the EP distribution in negative region, i.e., $P(\Delta S)$ for $\Delta S < 0$ using the extreme value theory. To perform this task, we consider a corresponding accumulated distribution of $P(\Delta S)$, i.e., $\tilde F(x)={\rm Prob}\{\Delta S \ge x\}$, that is, $\tilde F(x)=\int_x^{\infty} d\Delta S P(\Delta S)$ for $x < 0$. The probability that the smallest value $y_i$ of a set $i$ of size $\ell$ is larger than $x$ is given as $\tilde F^{\ell}(x)$, which is denoted as $\tilde G_\ell(x)$. Next, the probability that the smallest value among those $\{y_i\}$ ($i=1,\cdots, m$) elements is larger than $x$ is given as $\tilde G_\ell^m(x)$, which is equal to $\tilde F^{\ell m}(x)$. To have the asymptote in the same functional form, $\tilde G_{\ell}(x)$ and $\tilde G_{\ell}^m(x)$ need to have linearly shifted argument. That is, $\tilde G_\ell^m(x)=\tilde G_\ell(a_m x+b_m)$. We find that $a_m=1$ and $b_m=\tilde c\ln m$ ($\tilde c>0$). Therefore, we obtain the corresponding formula, 
\begin{equation}
\tilde F^\ell(x)\equiv \tilde G_\ell(x)=\exp(-\ell e^{x/{\tilde c}+\tilde r}),
\end{equation}
where $\tilde r$ is an irrelevant constant and $x$ is implicitly negative. Therefore, the asymptote of the entropy production in the negative region is determined as 
\begin{align}
\tilde P(\Delta S)=-\frac{d\tilde F^\ell(x)}{dx}\frac{1}{{\ell \tilde F^{\ell-1}(x)}}\Big|_{x=\Delta S}
=\frac{1}{\tilde c}\exp(-e^{x/{\tilde c}+{\tilde r}})e^{x/{\tilde c}+\tilde r}\Big|_{x=\Delta S}
\to {\rm exp}(\frac{1}{\tilde c}\Delta S).
\label{eq:11}
\end{align}
We remark that $\Delta S$ is negative. Thus Eq.~\eqref{eq:11} may be rewritten as 
\begin{align}
\tilde P(\Delta S)\to \exp(-\frac{1}{\tilde c}|\Delta S|)
\end{align} 
in the limit $\Delta S \to -\infty$. We find that the constant $\tilde c$ also depends on networks and has different values from $c$. The EP distribution is not symmetric. Numerically, we measure $1/\tilde c$ in Figs.~\ref{fig:gumbel_negative}(a), (b), (c), and (d), and list those values with $1/c$ as follows: $1/\tilde c$ ($1/c)=3.67 (2.52), 6.0 (4.47), 3.87 (2.88)$, and $3.87 (2.76)$ in Figs.~\ref{fig:gumbel}(b), \ref{fig:gumbel_all}(b), \ref{fig:gumbel_all}(d), and \ref{fig:gumbel_all}(f), respectively. The difference $1/\tilde c-1/c$ is roughly close to one for BA and CL scale-free networks, but deviates more than one for ER networks. For SF networks, we may say that the detailed FT shown in Eq.~\eqref{eq:2} is roughly satisfied, even though the numerical values of $1/\tilde c- 1/c$ is not exactly one. In fact, ER networks are too random in connections to serve as a state space.    

\section{Discussion}
Here we derive the asymptotes of a pure exponential distribution. Assume that $P(x)$ is a pure exponential function, that is, $P(x)=\frac{a}{2}\exp(-a|x|)$ for $-\infty < x < \infty$. Then, $F(x)\equiv 1-\int_x^{\infty} P(y)dy=1-\exp(-ax)/2$ for $x > 0$. For large $x$, $F(x)\approx \exp(-\frac{1}{2}e^{-ax})$. Thus, 
\begin{align}
F^{\ell}(x)\equiv G_\ell(x)\approx \exp(-\frac{1}{2}e^{-a(x-\frac{1}{a}\ln \ell)}). 
\end{align} 
Therefore, we conclude that $\alpha_\ell=1$ and $\beta_\ell=-(1/a)\ln \ell$. Therefore, we confirm that if $\alpha_m=1$ and $\beta_m\propto \ln m$ for the Gumbel distribution, the original distribution would follow a pure exponential function asymptotically. On the other hand, it was shown~\cite{wolpert} that the asymptote of a Gaussian distribution follows the Gumbel distribution, but with $\alpha_\ell \sim 1/\sqrt{\ln \ell}$ and $\beta_\ell\sim \sqrt{\ln \ell}$. We also put the relation between $a_m, b_m$ in $G^m_{l}(x) = G_{l}(a_m x + b_m )$  and $\alpha_l, \beta_l$, written as $ a_m =\alpha_{l}/ \alpha_{lm} $ and $b_m = \beta_{l} - a_m \beta_{lm}$. For the exponential form with $\alpha_l =1$ and $\beta_l \sim \ln l$, one can get $a_m =1$ and $b_m \sim \ln m$ while for the Gaussian, $a_m \sim \sqrt{ 1 + \ln m / \ln \ell} $ and $b_m \sim \ln m / \sqrt{ \ln \ell} $.

%Following the textbook,  we can define $G(x) = \lim_{l \to \infty} G_{l}( \alpha_l x + \beta_l )$ where $\alpha_l$ and $\beta_l$ are proper scaling parameters around Eq (3). For your information,  if we have the pure exponential function such as $exp[-ax]$, $\alpha_l = 1$ and $\beta_l = (1/a) \ln l$, and for the Gaussian function, $\alpha_l \sim 1/\sqrt{\ln l}$ and $\beta_l \sim \sqrt{\ln l}$. We also put the relation between $a_m, b_m$ in $G^m_{l}(x) = G_{l}(a_m x + b_m )$  and $\alpha_l, \beta_l$, written as $ a_m =\alpha_{l}/ \alpha_{lm} $ and $b_m = \beta_{l} - a_m \beta_{lm}$. For the exponential form with $\alpha_l =1$ and $\beta_l \sim \ln l$, one can get $a_m =1$ and $b_m \sim \ln m$ while for the Gaussian, $a_m \sim \sqrt{ 1 + \ln m / \ln \ell} $ and $b_m \sim \ln m / \sqrt{ \ln \ell} $. So we need to change the $a_m$ and $b_m$ for the Gaussian below Eq. (13). 

%ii) For the Gaussian distribution, $P(x)=\frac{1}{\sqrt{2\pi}}\exp(-x^2/2)$, $F(x)$ is defined $F(x)\equiv 1-\int_x^{\infty} P(y)dy=1-\int_{-\infty}^x P(y)dy=\frac{1}{2}[1-{\rm erf}(x/\sqrt{2})]$, where erf is the so-called error function.  as one can show that if $G(x)$ is the asymptote of a Gaussian distribution, then .  

\section{Summary}\label{sec:summary}

In this paper, we considered the EP distribution arising from the complexity of the shortest pathways from one node to another on complex networks. We showed explicitly that this EP distribution satisfies well-known FTs, i.e., the integral FT and the detailed FT. To obtain the result, we considered a data packet transport problem in which a packet travels back and forth between every pair of nodes along each of the shortest pathways. At a branching node along the way, a packet chooses one branch randomly. The effect of this random choice reflects stochastic process in dynamics in nonequilibrium systems. Owing to the complexity of the shortest pathways, the probabilities of taking a shortest pathway in one direction and corresponding reverse direction can be different, resulting in a nonzero EP. We calculated this difference explicitly and determined the functional form of the EP distribution in the large-EP limit in positive and negative regions using the extreme value statistics. The asymptotes of EP distribution follow the Gumbel distribution, which behaves as $P(\Delta S)\sim e^{-\frac{1}{c}\Delta S}$ in positive region and $P(-|\Delta S|)\sim e^{-\frac{1}{\tilde c}\Delta S}$ in negative region, where $c$ and $\tilde c$ are constants in positive and negative regions, respectively. The constants depend on networks and are different from each other. The numerical differences $1/\tilde c-1/c$ are roughly one when networks are scale-free, for which the detailed FT hold.   

In the stochastic thermodynamics, the fluctuation theorems were derived from the total EP that is based on the trajectory-dependent EP in stochastic process. However, because the trajectories of stochastic dynamic process are rather virtual, one can hardly imagine the origin of the total EP and understand the origin of the fluctuation theorems. In this study, one can identify the reverse trajectory easily and can calculate the total EP explicitly. Thus, our result would be pedagogically helpful not only for understanding the concept of trajectory-dependent EP in stochastic processes, but also for exploring nonequilibrium fluctuations and their relationship in complex networks.   

\section*{Acknowledgement}
This work was supported by the National Research Foundation of Korea by Grant No. 2014R1A3A2069005 (BK), 2017R1D1A1B03030872 (JU), 2017R1A6A3A11033971 (DL), the Brain Pool program by Grant No. 171S-1-2-1882 (YWK), and the IBS by Grant No. IBS-R020-D1 (HKP). BK thanks C. Kwon and J. D. Noh for useful discussion. 

\appendix
\section{Network models}
\begin{itemize}
\item[i)] A BA network is constructed as follows: At the beginning, there exist $m_0$ nodes in the system. At each time step a node is added with $m$ links in the system, where $m \le m_0$. Each link is connected to a node $i$ with degree $k_i$ with the probability $p(k_i)=k_i/\sum_j k_j$. This process is repeated until the total number of nodes in the system becomes $N$. This model generates a scale-free network with the degree distribution $P_d(k)\sim k^{-\gamma}$ with $\gamma=3$~\cite{ba}.
\item[ii)] A CL network is constructed as follows: At the beginning, there exist a fixed number of $N$ nodes indexed $i=1,\dots,N$ in the system. Then, a node $i$ is assigned a weight of $w_i=(i+i_0-1)^{-\mu}$, where $\mu \in [0,1)$ is a control parameter, and $i_0\propto N^{1-1/2\mu}$ for $1/2 < \mu < 1$ and $i_0=1$ for $\mu < 1/2$. Then, two different nodes $(i, j)$ are selected with their probabilities equal to the normalized weights, $w_i/\sum_k w_k$ and $w_j/\sum_k w_k$, respectively, and a link is added between them unless one already exists. This process is repeated until $pN$ links are created in the system, where $p$ is a control parameter. There exists a percolation threshold $p_c$, above which a macroscopic-scale large cluster is generated. We considered the data packet transport problem on such large networks. The obtained network is scale-free in degree distribution with the exponent $\lambda=1+1/\mu$.
\item[iii)] An Erd\H{o}s-R\'enyi network is constructed as follows:  At the beginning, there exist a fixed number of $N$ vertices in the system. At each time step, two nodes are selected randomly. They are connected with a link unless they are already connected. This process is repeated until $pN$ links are created in the system, where $p$ is a control parameter. It is known that $p_c=1/2$ is the percolation threshold. Thus,  for $p > p_c$, a macroscopic-scale large cluster is generated. We considered the data packet transport problem on such large networks. The obtained network has the degree distribution following a Poisson distribution.   
\end{itemize}

\end{document}